\documentclass{iaus}
\usepackage{graphicx}

\title[Concluding remarks] 
{Concluding remarks}

\author[Jean-Paul Zahn] 
{Jean-Paul Zahn}

\affiliation{LUTH, Observatoire de Paris, 92195 Meudon, France \break email: Jean-Paul.Zahn@obspm.fr}
\pubyear{2006}
\volume{239}
\pagerange{119--126}
\date{?? and in revised form ??}
\jname{Convection in astrophysics}
\editors{F. Kupka, I. W. Roxburgh \& K. L. Chan, eds.}
\begin{document}

\maketitle

\firstsection 
\section{From Nice to Prague}

Thermal convection occurs in most objects that populate our Universe, whenever radiation is insufficient to transport the heat because the medium is too opaque.  In astrophysical objects convection involves a wide range of spatial and temporal scales - experts call this turbulence - which makes it rather difficult to model. For this reason convection remains one of the major uncertainties when modeling stars and planets, and this is partly true also for accretion disks\footnote{In fact, H. Klahr showed us that convection plays little role in accretion disks, because it is not a self-sustained process.}. However, substantial progress has been achieved during the past years, both in the numerical simulation of convective regions and in the observation of convective flows by various new techniques. 

There was thus a need to confront and to discuss the most recent results, and that is why Friedrich Kupka and Ian Roxburgh organized this symposium, exactly 30 years after the IAU colloquium 38 which was hosted by the Nice observatory in the frame of the XVIth General Assembly. That colloquium  was then more in the style of a workshop, gathering 35 participants (of whom a few are here today), whereas as many as 300 persons registered in the present symposium.

In his introductory talk,  Ed Spiegel anticipated that the 1976 meeting would be a confrontation between ``those who want to write down the full equations and solve them, who have virtue but no results that apply directly to stars'' and ``those want to write down an algorithm for computing stellar structure that contains adjustable parameters which can be fit to well known cases''. Only a few did then believe in the first approach; I remember in particular  Eric Graham who presented there the first 3-D calculation of stratified convection, and \AA ke Nordlund who during the coffee breaks showed us his early 2-D simulations in form of a thumb-nail movie. But at present it appears that the ``full equations solvers'' are going to win the game, thanks to the ever increasing performances of massively parallel supercomputers. Indeed, 3-dimensional simulations have now reached a high level of realism, which was demonstrated again and again during this symposium. F. Cattaneo reminded us of the main steps of that development, and of its major actors, of whom a few are still active, as attested by their participation in the present meeting.

\section{The triumph of 3-D simulations}
Among the myriad of objects that host a convective region, it is the Sun that, quite naturally, draws most attention.
Of course, no single numerical code is able to encompass the whole convection zone of solar-type stars, because it extends over too many scale-heights. That is why some codes are designed to describe the upper layers of stars, including the photosphere: they work in Cartesian geometry and include explicitly the transfer of radiation, sometimes treating it in non-local thermal equilibrium. Their great success is to render  the solar granulation in exquisite detail, once the seeing effects have been taken in account. Moreover, these local codes reproduce with high accuracy the observed line profiles, as illustrated once more by M. Asplund and M. Steffen during this symposium,
and they dispense with the ad-hoc introduction of the so-called micro-turbulence, with which stellar atmospherists have learned to live. It should be added that with A-type stars the agreement is somewhat less perfect, in particular concerning the strong lines [Kochukov], which proves, if it were necessary, that in these stars convection differs appreciably from that in the Sun [Robinson].

Other codes are tailored for deeper convection and some operate on global scale and in spherical geometry. They now succeed in reproducing correctly the differential rotation of the solar convection zone, as it was revealed through helioseismology, with the equator rotating faster than the poles and the angular velocity remaining almost constant with depth.  More and more, such global codes are being applied to other stars: to brown dwarfs where the formation of dust must be taken in account [Mohanty; Ludwig; Helling], to pre-supernovae where they include nuclear burning [Arnett], to giant stars in which the convection zone extends over multiple scale heights [Palacios], to A-type-stars where convection occurs in the core (Brun et al. 2005), and even to giant planets [Chan], etc. 

Both types of codes, the local and the global ones, display a striking property of stratified convection, which was already hinted in the crude 1-D calculations we performed with Ed Spiegel in the seventies, namely the asymmetry between the slow up-welling flows that occupy most of the domain and the down-going motions, which are organized in fast narrow plumes. These plumes penetrate into the stable region beneath, where they induce mixing and generate internal gravity waves. Just how far they penetrate is still an unsettled issue, because the spatial resolution presently available is insufficient to account for the stiffness of the stable stratification that is achieved in a star [Brummell]. In other words the thermal diffusivity, which must be enhanced to meet numerical requirements, erodes the 
 (anti-)buoyancy force which is supposed to brake the downdrafts, thus predicting deeper penetration than does actually occur. Semi-convection too is still an unsolved problem, though G. Bascoul reported some attempts to model it that are inspired by the treatment of thermohaline convection.

These 3-D codes are now being used for various other purposes, as was described during this meeting: to compute the excitation of acoustic modes [Stein; Samadi et al.], to validate local helioseismology for mapping the deep convective flows [Hanasoge], to analyze in detail the fine properties of convection [Rincon], and to test closure models [Belkacem et al.; Kupka]. In the same spirit, K. Penev et al. (2006) take such a code to study the damping of an oscillating large-scale shearing flow; not only do they validate the use of a turbulent viscosity for this problem - this is good news to those who want to couple pulsation with convection - but they settle the long-standing issue of how the efficiency of tidal dissipation is reduced in close binary stars when the period of the tide becomes shorter than the convective turn-over time.
\section{How to lower the cost -- what else can be done?}
However these 3-D simulations suffer from a serious drawback: they are much too heavy tools to build models for the purpose of describing stellar evolution, except when one deals with very fast phases [cf. Eggleton].  That is why the crude mixing-length approach is still so widely used. It has even been called (jokingly) a ``blessing'' by one of the speakers, to the dismay of those who know all about its shortcomings. 

What else can one do? 
A few, led by D. R.  Xiong and V. Canuto, still believe  in a 1-D approach, but they strive boldly to go beyond the crude mixing-length recipe. For years now they have been developing a treatment that is inhomogeneous in the horizontal directions and non-local in the vertical one. This procedure involves various closure approximations that need to be justified by comparison with laboratory experiments, geophysical measurements or high-resolution calculations. F. Kupka  has spent much effort in comparing such Reynolds stress prescriptions with 2 and 3-D simulations, but so far the results have been somewhat disappointing. Nevertheless, he and Canuto are convinced that the situation will improve by ``plumenizing'' their model, i.e. by making better account of the plumes, which advect heat rather than diffusing it. Similar prescriptions have been applied to study the coupling between convection and pulsation in red giants [Xiong] and in $\delta$ Scuti stars [Montalban].

One way to incorporate in stellar modeling the benefits of 3-D radiation-hydrodynamic simulations would be to calculate a grid of horizontally averaged envelope models covering the Hertzsprung-Russel diagram, much as N. Baker and S. Temesvary did with 1-D models in the early sixties, when the computer resources were still scarce. This was undertaken some time ago by H.-G. Ludwig and his colleagues, with a 2-D code, and it would be highly desirable to repeat that exercise with more realistic 3-D simulations. But in the discussion some expressed the opinion  that such an enterprise would be judged too costly in manpower, and not enough rewarding:  Baker and Temesvary's tables were published by NASA, broadly  distributed and widely used, but they are not even referenced in ADS.

\section{A powerful tool to determine surface abundances}

A much discussed problem during this symposium was that raised by the new abundance determinations of carbon, nitrogen and oxygen in the Sun, which were carried out by M. Asplund and his collaborators [see also M. Steffen]. By fitting the observed line profiles to those predicted with their 3-D radiation-hydrodynamic code, they found that these elements are up to 2 times less abundant than previously determined through classical model atmosphere calculations. Part of the correction is due also to better atomic data and to a non-local thermal equilibrium treatment. This revised metallicity puts the Sun to a level comparable to that of other solar-type stars in its neighborhood, whereas before it appeared somewhat as an exception. But that lower metallicity implies a reduced opacity in the solar interior, and hence a smaller depth of the convection zone, which is in conflict with the depth inferred from helioseismology, at least when using standard reference models. What is to blame? Having listened to all arguments, it seems to me that the new abundances are indeed trustworthy. Perhaps the problem lies, as John Bahcall was already claiming, with the abundance of neon, a major contributor to the opacity below the convection zone: it is true that its abundance is not derived directly from line profiles but deduced from the Ne/O ratio observed in the solar corona, where some element separation may occur. Another possibility is that the solar reference model which is used for such comparisons is inadequate because it ignores convective penetration.

\section{The effects of convection are not confined to convection zones}
\label{gw}

Although the symposium was dedicated to convection, there was much talk also about mixing in radiation zones. Such mixing is revealed by the appearance, on the surface of stars, of elements that have been synthesized in the deep interior  [Drake, Smilianic, Spite, Tautvaisiene, Tsuji], or by the depletion of fragile elements, such as lithium, which are destroyed through nuclear burning. Moreover, microscopic diffusion (plus radiative levitation and gravitational settling) is clearly impeded by some macroscopic mixing process, otherwise many elements would appear either extremely overabundant or underabundant. Convective penetration and overshoot may be responsible for part of this mixing, but in most cases a deeper mixing is required to explain the observations. One plausible cause is the mixing due to rotation, either through  a large-scale meridional circulation or through turbulence produced by the shear of differential rotation. Such rotational mixing accounts rather well for the properties of massive stars, which are rapid rotators. But it cannot explain alone the uniformly rotating radiative interior of the Sun, which requires a much more efficient  process to transport angular momentum. Magnetic stresses have been invoked, but the cyclic dynamo field cannot penetrate deep enough, due to the skin effect, and a fossil field would imprint on the interior the differential rotation of the convection zone
(Brun \& Zahn 2006). Another possibility is the transport of angular momentum by long-period gravity waves emitted at the base of the convection zone.
It was recently shown by C. Charbonnel and S. Talon (2005) that such waves are indeed able to extract angular momentum from the solar interior as the Sun is gradually spun down by losing matter through its magnetized wind. When applied to old metal-poor stars, this mechanism produces exactly the depletion of lithium that is needed to reconcile the observed abundance with that produced by the Big Bang, according to the results of WMAP. These internal gravity waves seem also to play an important role in other objects, such as giant planets [Glatzmaier] or stars in their advanced stages [Arnett].

\section{Towards a realistic model of the solar dynamo}

But for many solar and stellar physicists, the most spectacular manifestation of convection is magnetic activity. Convection is responsible for the magnetic field that is observed in late-type stars, and much effort has gone into explaining how this field is produced. As we were reminded by S. Brun, the current paradigm is the so-called alpha-omega dynamo first proposed by E. Parker, where a poloidal field is sheared into a toroidal field in the tachocline, the toroidal field thereafter being twisted somewhere above into a poloidal field, thus closing the loop. Numerical simulations easily generate a random field of substantial strength, both in convective shells and in convective cores, 
and such a field certainly exists in the Sun [cf. Trujillo-Bueno]. But so far it has not been possible to render the cyclic behavior observed in the Sun and many other stars, presumably because the computational domain was restricted to the unstable region. Fortunately the situation is improving since the stable tachocline has been included, and the first results obtained by M. Browning et al. (2006) show that the field seems to conform there to the alpha-omega scheme. It remains to explain how a fully convective star, contrary to all expectations,  can host an almost dipolar field, as was recently revealed by J.-F. Donati et al. (2006) using the powerful spectropolarimeter ESPaDOnS at CFHT. And the magnetic fields of the giant planets have not ceased to puzzle us, as we were reminded by F. Busse, who has long been tackling the problem
[see also Christensen].

\section{The moment of truth: facing observational tests}
So far I have been describing only theoretical advances, but these have been achieved mainly because observations of various types impose more and more stringent constraints. Helioseismology is a wonderful tool, as we all know, and this was demonstrated again by Alan Title in his brilliant invited discourse.  (I cannot resist to recall that it was in Nice also, one year before the convection colloquium, that Franz Deubner presented his first results on the acoustic modes he had observed at the surface of the Sun.) Without helioseismology, as we were reminded by J. Christensen-Dalsgaard, we wouldn't know the depth of the solar convection zone, there would be no hint of a tachocline, we would still believe that the convection zone is rotating on cylinders and that the solar core is spinning faster than the surface. Furthermore, local seismology is now able to deliver maps of the convective flows below the surface [Kosovichev, Stein]. 
Thanks to adaptive optics, images of the solar surface are routinely obtained at the diffraction limit. Moreover, convection on the surface of other stars is increasingly constrained by high-resolution spectroscopy, as was demonstrated by J. Landstreet [cf. also Cauzzi, Kochukov, Monier, Smalley]. 

Asteroseismology looks as promising as helioseismology, although the mode identification is much more difficult with stars that rotate substantially faster than the Sun.  This powerful technique is already providing most interesting results, from ground and from space [Kjeldsen; Straka et al.]. And much more is to come. Before these lines go into print, the COROT satellite will have been launched, and no doubt it will have delivered many surprises.

\section{On my wish-list}
The privilege of someone who is asked to draw the conclusions of such a meeting is to be given the opportunity to mention the problems that he or she would like having solved in priority. Here is my list.
\begin{itemize}

\item On the top I would rank the precise and reliable prediction of the entropy jump in the solar convection zone, using the most advanced 3-D simulations. Until this is straightened out, we will be obliged to calibrate somehow this jump, or equivalently the mixing-length if we are using that treatment. And we will thus depend on other data, such as isotope ratios measured in meteorites, to determine the age of the Sun.

\item Next on my list is resolving the conflict between the depth of the solar convection zone inferred from acoustic sounding and that which is predicted by the new CNO abundances, these being deduced from the comparison of spectral lines profiles that are observed with those which are obtained through 3-D simulations. Which is to blame for this discrepancy? An underestimate of the neon abundance? The fact that convective penetration is not taken in account in the reference model?

\item Increase the resolution of the simulations of penetrative convection. No one went as far as Brummell et al. (2002) with a mesh of $512\times 512\times 575$, and even these calculations are still too diffusive to enforce a nearly adiabatic stratification in the stable region, as was pointed out by Rempel (2004). The role of rotation need also to be elucidated and quantified [Brummell; K\"apil\"a].

\item Use such  high-resolution simulations of convective penetration to predict the flux and spectrum of the long period internal gravity waves that are emitted at the base of the convection zone of solar-type stars. This will allow to check whether such waves can, as it is claimed by Charbonnel and Talon (2005), extract the angular momentum from the solar core as the Sun is gradually spun down. As we mentioned above in \S\ref{gw}, this has far reaching consequences, since it governs the Li abundance at the surface of the oldest stars, which provides one of the rare constraints on the Big Bang theory.

\item Likewise, with the codes written to model surface convection one should predict the spectrum and the amplitude of the gravity modes excited near the top of the solar convection zone, to verify whether one can hope to detect these modes with the helioseismic instruments that are currently in use, or planed for the near future. The only estimate available so far is that by Goldreich et al. (1994), but it involves several simplifying assumptions about turbulent convection that are not justified.

\item Finally, it goes without saying that we all wait eagerly for a realistic model of the solar dynamo, displaying cyclic reversals of the magnetic field.

\end {itemize}

\medskip

For some of these problems the solution is in reach, but for others we will probably have to wait until the next symposium on astrophysical convection!
\medskip

May I conclude by declaring how much I enjoyed this most interesting symposium, and by expressing our thanks to  Friedrich and Ian for having brought us together in this wonderful city of Prague.


\begin{thebibliography}{}

\bibitem[]{}NB. Quotations in [\,.\,.\,.\,] refer to contributions or posters presented at this symposium.

\medskip

\bibitem[]{}
{{Browning}, M.K., {Miesch}, M.S., {Brun}, A.S., {Toomre}, J.} 2006,
  \textit{ApJ}, 648, L157
  
  \bibitem[]{}
{{Brummell}, N.H., {Clune}, T.L., {Brun}, A.S., {Toomre}, J.} 2002,
  \textit{ApJ}, 570, 825
  
  \bibitem[]{}
{{Brun}, A.S., {Browning}, M.K., {Toomre}, J.} 2006,
  \textit{ApJ}, 629, 461
  
  \bibitem[]{}
{{Brun}, A.S., {Zahn}, J.-P} 2006,
  \textit{A\&A}, 457, 665


 \bibitem[]{}
  {{Charbonnel}, C., {Talon}, S.} 2005,   \textit{Science}, 309, 2189

  \bibitem[]{}
  {{Donati}, J.-F., {Forveille}, T., {Cameron}, A.C.,
	{Barnes}, J.R., {Delfosse}, X., {Jardine}, M.M., 
	{Valenti}, J.A.} 2006,   \textit{Science}, 311, 633
  
    \bibitem[]{}
{{Goldreich}, P., {Murray}, N., {Kumar}}  1994,
  \textit{ApJ}, 424, 466
  
  \bibitem[]{}
     {Penev, K., Sasselov, D., Robinson, F., Demarque, P.} 2006,  
    \textit{ApJ} (in press) astro-ph/0607016
    
  \bibitem[]{}
{{Rempel}, M.} 2004,
  \textit{ApJ}, 607, 1046

  \bibitem[]{}
     {Spiegel, E.A. \& Zahn, J.-P.} 1977,
     \textit{Problems of stellar convection: Proc. IAU Coll. 38},
     Lecture Notes in Physics (Heidelberg: Springer), vol.\ 71


 
\end{thebibliography}
\end{document}